\documentstyle[preprint,aps]{revtex}
\date{\today}
\title{Effects of screening on the Hofstadter butterfly}
\author{Vidar Gudmundsson}
\address{Science Institute, University of Iceland, Dunhaga 3,
         IS-107 Reykjavik, Iceland.}
\author{Rolf R. Gerhardts}
\address{Max-Planck-Institut f\"ur Festk\"orperforschung,
         Heisenbergstra{\ss}e 1,
         D-70569 Stuttgart,\\  Federal Republic of Germany}
\tighten
\begin{document}
\draft
\maketitle
\abstract{We study, within the Hartree approximation, the effects of the
          electron-electron interaction on the energy spectrum
          of a two-dimensional electron gas in a perpendicular
          homogeneous magnetic field and a lateral
          superlattice potential with square symmetry.
          Due to the strong screening effects,
          the bandwidth of the Landau bands oscillates strongly
          with their filling. For short enough
          periods and strong enough modulation of the superlattice
          potential, the miniband structure of the Landau bands
          can be resolved in the thermodynamic density of states.
}

\section{Introduction}
The study of the characteristics of a two-dimensional electron gas (2DEG)
in a homogeneous perpendicular quantizing magnetic field $\vec B$
and a lateral periodic superlattice potential has been revived, since due
to recent technological developments it now is possible to  manifacture such
systems
with controlable properties in semiconductor heterostructures.
Measurements of the magnetotransport
in superlattices weakly modulated in one or two directions have
revealed novel oscillations in the conductivity, the Weiss oscillations,
reflecting the commensurability of the cyclotron radius
$R_c=l^2k_F=v_F/\omega_c$ of the electrons at the Fermi energy
$E_F=\hbar^2k^2_F/(2m^*)$ and the modulation
period $L$, $l=(c\hbar /eB)^{1/2}$ is the magnetic
length.\cite{Gerhardts89:1173,Weiss89:179}
The Weiss oscillations are superimposed on top of the Shubnikov-de Haas
oscillations which are well known in homogeneous 2DEG's and caused by the
commensurability of the Fermi wavelength $\lambda_F$ and $l$.

Transport calculations have been able to explain the observed difference
in the Weiss oscillatons for weakly 1D and 2D modulated superlattices
in a strong magnetic field.\cite{Pfannkuche92:12606}
The periodic potential lifts the degeneracy of the Landau levels (LL's)
leading to Landau bands (LB's) of oscillating width periodic in $B^{-1}$.
The bandwidth is
%% FOLLOWING LINE CANNOT BE BROKEN BEFORE 80 CHAR
determined\cite{Gerhardts90:1473,Zhang90:12850,Winkler89:1177,Beenakker89:2020,Vasilopoulos89:2120}
by the commensurability of $R_c$ and $L$.

The single-particle energy spectrum for
this problem has been investigated by several
%% FOLLOWING LINE CANNOT BE BROKEN BEFORE 80 CHAR
researchers\cite{Harper55:874,Azbel64:634,Langbein69:633,Rauh75:K9,Hofstadter76:2239,Thouless87:101}
culminating in Hofstadter's butterfly,\cite{Hofstadter76:2239} a graph
showing the complicated self-similar splitting of a LB into minibands
as a function of the magnetic flux through a unit cell of the lattice.
The effects of the splitting of LB's into minibands has,
to the best of our knowledge, not been observed directly  in the currently
common superlattices ($L\sim 300\: $nm $n_s=3\times 10^{11}\: $cm$^{-2}$)
since usually several LB's are occupied, which may overlap near $E_F$.
Transport calculations based on the Hofstadter energy spectrum and taking into
account short-range impurity scattering have, however, succeeded in recovering
all the observed features of the magnetotransport coefficients for
large-period
superlattices.\cite{Pfannkuche92:12606} In the limit of a very high
magnetic field the intricate gap structure of the Hofstadter butterfly
is robust but long-range disorder tends to smear out the fine structure in
the density of states.\cite{Wulf93:6566}

The Hofstadter butterfly is obtained in two complimentary
but mathematically equivalent limits:\cite{Langbein69:633}
first, in a strong lattice potential and a weak magnetic field
in the tight-binding
method,\cite{Harper55:874,Langbein69:633,Hofstadter76:2239} and,
second, a weak periodic perturbation of Landau quantized
2DEG.\cite{Langbein69:633,Thouless87:101,Usov88:2565}
Both approaches deliver the so-called Harper's
equation\cite{Harper55:874,Langbein69:633} determining the energy spectrum.
The energy spectrum has been investigated in the intermediate
region in the absence of collision broadening. The coupling of the LL's
by the external periodic potential
strongly reduces the original high symmetry of the
Hofstadter spectrum but retains a very complicated
subband structure.\cite{Kuhn93:8225,Petschel93:239}
It has been possible to transform the equation of motion, the
Schr{\"o}dinger equation, into a vectorial form reminiscent
of Harper's equation, but exact under the most general conditions
and exhibiting chaos in the classical limit.\cite{Petschel93:239}

All the above-mentioned theoretical investigations of the energy spectrum
of a 2DEG in a periodic potential and a homogeneous perpendicular
magnetic field have neglected the effects of the electron-electron
interaction. In the present publication we initiate, within the Hartree
approximation,  the study of these effects on the electronic energy spectrum
and the thermodynamic density of states.

\section{Model}
To describe the electrons in the conduction band of a lateral
superlattice at the AlGaAs-GaAs interface in a constant perpendicular
magnetic field $\vec B=B\hat z$ we employ a model of strictly
two-dimensional electron gas (2DEG), with the three-dimensional charge density
given by $-e n_s(\vec{r})\delta(z)$, and $\vec{r}=(x,y)$. The orthogonal
superlattice is spanned by the lattice vectors
\begin{equation}
      \vec R=m\vec l_1 + n\vec l_2, \quad n,m\in Z.
\end{equation}
$\vec l_1=l_1\hat x$, and $\vec l_2=l_2\hat y$ are the primitive
translations of the Bravais lattice ${\cal B}$.
The corresponding reciprocal lattice ${\cal R}$ is spanned by
\begin{equation}
      \vec G=G_1\vec g_1 + G_2\vec g_2, \quad G_1, G_2 \in Z,
      \label{vecG}
\end{equation}
where
\begin{equation}
      \vec g_1=2\pi\frac{\hat x}{l_1}, \quad \vec g_2=2\pi\frac{\hat y}{l_2}
\end{equation}
The external periodic potential the electrons are moving in
is taken to be of the simple form
\begin{equation}
      V(\vec r)=V_x\cos (g_1x) + V_y\cos (g_2y).
      \label{V}
\end{equation}
The electron-electron interaction is included in the Hartree approximation
(HA) leading to the effective single-electron Hamiltonian
\begin{equation}
      H=H_0+V_H(\vec r)+V(\vec r),
      \label{H}
\end{equation}
where $V_H(\vec r)$ is the effective potential in a medium with
a dielectric constant $\kappa$,
\begin{equation}
      V_H(\vec r)=\frac{e^2}{\kappa}\int_{R^2} d^2r' ~
      \frac{n_s(\vec r\: ')-n_b}{|\vec r-\vec r\: '|} ~~,
      \label{VH}
\end{equation}
felt be each electron and caused by the total charge density of the
2DEG, $-en_s(\vec r)$, and the neutralizing background charge density
$+en_b=+e\langle n_s(\vec r)\rangle$.

The periodic external potential $V(\vec r)$ and the constant external
magnetic field imply that all physical quantities of the
noninteracting system are periodic with respect to translations of
$\vec R\in {\cal B}$. Since the Coulomb interaction is
given by a central potential between pairs of electrons,
the periodicity of the system is not broken by the electron-electron
interaction. The periodicity of the Hartree potential (\ref{VH})
follows from that of $n_s(\vec r)$.

The translation operator
\begin{equation}
      T(\vec R)=\exp \left( -\frac{i}{\hbar}\vec R\cdot\vec p \right) ,
      \label{T}
\end{equation}
where $\vec p$ is the momentum canonical to $\vec r$,
does not commute with $H_0$, $[T(\vec R), H_0]\neq 0$. Due to the
presence of the vector potential $\vec A$ in $H_0$,
\begin{equation}
      H_0=\frac{1}{2m}\left( \vec p +\frac{e}{c}\vec A \right) ^2,
      \quad \vec B =\vec\nabla\times\vec A
      \label{H0}
\end{equation}
an additional
gauge transformation is necessary. Defining the magnetotranslation
\begin{equation}
      S(\vec R)=\exp \left( \frac{ie}{\hbar c}\chi \right) T(\vec R),
      \label{MT}
\end{equation}
where $\chi$ is obtained from $T^{-1}\vec A T=\vec A +\vec\nabla\chi$,
we have for the mechanical momentum $\vec\pi =\vec p + (e/c)\vec A$
and the position $\vec r$
\begin{eqnarray}
      S^{-1}(\vec R)\vec r  S(\vec R) &=& \vec r+\vec R \nonumber \\
      S^{-1}(\vec R)\vec\pi S(\vec R) &=& \vec\pi .
\end{eqnarray}
The Hamiltonian $H$ is thus invariant with respect
to the magnetotranslations, $[S(\vec R), H]=0$.

$H_0$ defines the natural length and frequency scales for the system
by the magnetic length $l$ and the cyclotron frequency $\omega_c$
\begin{equation}
      l=\sqrt{\frac{c\hbar}{eB}}, \quad \omega_c=\frac{eB}{mc}.
\end{equation}
Generally the magnetotranslations (\ref{MT}) do not commute,
$[S(\vec R_1), S(\vec R_2)]\neq 0$, expressing the fact that $l$
is not commensurate with the lattice. But if an integer number of
flux quanta $\Phi_0=(hc/e)$ flows through the lattice unit cell
$\vec l_1 \times \vec l_2$, then $[S(\vec R_1), S(\vec R_2)]=0$
and the magnetic field assumes the values
\begin{equation}
      B=g_l\frac{\Phi_0}{|\vec l_1\times\vec l_2|}, \quad g_l\in Z.
\end{equation}
For these values of $B$ the magnetotranslations (\ref{MT})
of the lattice vectors commute with each other, with $H$, and also with $H_0$.
Thus, the known eigenvectors of $H_0$ can be chosen to be also eigenvectors of
the unitary magnetotranslations with simple eigenvalues
\begin{eqnarray}
      S(\vec l_1)\psi &=& e^{i\theta_1}\psi \nonumber \\
      S(\vec l_2)\psi &=& e^{i\theta_2}\psi, \quad \theta_i\in [-\pi,\pi ].
\end{eqnarray}
The above eigenequations can be reinterpreted as periodic boundary
conditions for the wave functions within one unit cell of ${\cal B}$,
and the eigenfunctions form a complete basis for each pair
of phase angles $(\theta_1, \theta_2)$.

Here we use the symmetric basis functions constructed by
Ferrari\cite{Ferrari90:4598} and used by
Silberbauer.\cite{Silberbauer92:7355,Silberbauer94:PhD}
A finer lattice with only one flux quantum
through a unit cell is defined by the primitive vectors
\begin{equation}
      \vec c=\frac{\vec l_1}{p}, \quad \vec d=\frac{\vec l_2}{q}, \quad
      pq=g_l, p,q\in N.
      \label{cd}
\end{equation}
With the symmetric gauge $\vec A=(B/2)(-y,x)$ the functions
\begin{equation}
      \phi^{\mu\nu}_{n_l}(\vec r)=\frac{1}{\sqrt{pq}}
      \sum^{\infty}_{m,n=-\infty}\left[ S(\vec c)e^{-i\mu}\right] ^m
      \left[ S(\vec d)e^{-i\nu}\right] ^n \phi_{n_l}(\vec r),
      \label{Ferrari}
\end{equation}
where
\begin{eqnarray}
      \mu &=& \frac{\theta_1+2\pi n_1}{p},
      \qquad n_1\in I_1=\{0,\dots p-1\} \nonumber \\
      \nu &=& \frac{\theta_2+2\pi n_2}{q},
      \qquad n_2\in I_2=\{0,\dots q-1\} \nonumber \\
      \phi_{n_l}(\vec r) &=& \frac{1}{\sqrt{2\pi n_l!l^2}}
      \left( \frac{x+iy}{\sqrt{2}l}\right) ^{n_l}
      \exp \left( -\frac{r^2}{4l^2} \right) , \quad n_l=0,1,2\dots,
\end{eqnarray}
form a complete orthogonal basis in the Hilbert space
${\cal H}_{\theta_1\theta_2}$ if $(\mu ,\nu )\neq (\pi ,\pi )$
for all $(n_1, n_2)\in I_1\times I_2$.
The norm of the $ \phi^{\mu\nu}_{n_l}$ can be shown to be
\begin{equation}
      ||\phi^{\mu\nu}_{n_l}||^2=\sum^{\infty}_{m,n=-\infty}(-1)^{mn}
      e^{i(\mu m+\nu n)}
      \exp \left(-\frac{1}{4l^2}|n\vec c+m\vec d|^2 \right) ~,
\end{equation}
and they have on the finer lattice the Bloch-type property
\begin{equation}
      S(\vec c)\phi^{\mu\nu}_{n_l}=e^{i\mu}\phi^{\mu\nu}_{n_l}, \quad
      S(\vec d)\phi^{\mu\nu}_{n_l}=e^{i\nu}\phi^{\mu\nu}_{n_l}.
\end{equation}
In order to utilize the Ferrari-basis (\ref{Ferrari})
$\{ \phi_{\alpha}\}$, each element of which satisfies
\begin{equation}
      H_0\phi_{\alpha}=E_{\alpha}\phi_{\alpha}, \quad \alpha =(\mu ,\nu, n_l),
\end{equation}
to find the energy spectrum $\{ \epsilon_{\alpha} \}$ and the eigen functions
$\{ \psi_{\alpha} \}$ of $H$,
\begin{equation}
      H\psi_{\alpha}=(H_0+V+V_H)\psi_{\alpha}=\epsilon_{\alpha}\psi_{\alpha},
\end{equation}
we need the matrix elements of the external periodic potential $V(\vec r)$
and the Hartree potential (\ref{VH}) in that basis.
Both $V(\vec r)$ and $V_H(\vec r)$
are periodic with respect to $\vec R\in {\cal B}$ and can be expressed
in terms of a Fourier series
\begin{equation}
      V(\vec r)=\sum_{\vec G\in {\cal R}}V(\vec G)e^{i\vec G\cdot \vec r},
\end{equation}
with the Fourier coefficients given by
\begin{equation}
      V(\vec G)=\frac{1}{A}\int_A d^2r\: V(\vec r)e^{-i\vec r\cdot\vec G}.
\end{equation}
Here $A$ stands both for the unit cell in ${\cal B}$ and it's area
$|\vec l_1\times\vec l_2|$. The task of deriving the matrix elements
of the external potentials is thus reduced to finding the matrix element
of the Fourier phase factor. This has been accomplished by Silberbauer
showing that the matrix elements can be expressed as rapidly converging
series of simple analytic
functions.\cite{Silberbauer92:7355,Silberbauer94:PhD}
The result is reproduced here, correcting
a simple typographical error in one of the sources\cite{Silberbauer92:7355}
\begin{equation}
      \langle n'_1n'_2n'_l|e^{i\vec G\cdot\vec r}|n_1n_2n_l\rangle
      =G^{n'_ln_l}(G)T^{n'_1n'_2}_{n_1n_2}(G)
      \frac{e^{-\frac{|G|^2l^2}{4}}}{||\phi^{n'_1n'_2}_{n'_l}||
      \: ||\phi^{n_1n_2}_{n_l}||},
\label{matrix-elements}
\end{equation}
where
\begin{equation}
      T^{n'_1n'_2}_{n_1n_2}(G)=\sum^{\infty}_{\Lambda\Omega =-\infty}
      (-1)^{\Lambda\Omega}e^{(\mu '\Lambda+\nu '\Omega)}
      \exp \left\{ -\frac{i}{2}G(\Lambda c+\Omega d)^*
      -\frac{|\Lambda c+\Omega d|^2}{4l^2}\right\}
\end{equation}
and
\begin{equation}
      G^{n'_ln_l}(G)=\left\{
      \begin{array}{cc}
            \sqrt{\frac{n_l}{n'_l}}e^{-\frac{|G|^2l^2}{4}}
            \left( \frac{iG^*l}{\sqrt{2}}\right) ^{n'_l-n_l}
            L^{n'_l-n_l}_{n_l}\left( \frac{|G|^2l^2}{2} \right)
            & \quad\mbox{if} \quad n'_l\geq n_l\\
            \sqrt{\frac{n'_l}{n_l}}e^{-\frac{|G|^2l^2}{4}}
            \left( \frac{iGl}{\sqrt{2}}\right) ^{n_l-n'_l}
            L^{n_l-n'_l}_{n'_l}\left( \frac{|G|^2l^2}{2} \right)
            & \quad\mbox{if} \quad n_l\geq n'_l
      \end{array}
      \right.
\end{equation}
Here a complex notation, $G=G_1|\vec g_1|+iG_2|\vec g_2|$
etc., has been used for the vectors
$\vec{G}$, Eq. (\ref{vecG}), and similarly for
$\vec c$, and $\vec d$, Eq. (\ref{cd}). $L^n_m$  are the associate Laguerre
polynomials.
Furthermore, the matrix elements $T^{n'_1n'_2}_{n_1n_2}(G)$ vanish
unless $G_1=n'_1-n_1-Mp$ and $G_2=n'_2-n_2-Nq$ for some $M,N\in Z$.

The Fourier coefficients for the Hartree potential are
\begin{equation}
      V_H(\vec G)=\left\{
      \begin{array}{cc}
            2\pi\frac{e^2}{\kappa |\vec G|}\Delta n_s(\vec G)
            & \quad \mbox{if} \quad \vec G\neq 0 \\
            0 & \quad \mbox{if} \quad \vec G=0
      \end{array}
      \right.  ,
\label{V_H}
\end{equation}
where
\begin{equation}
      \Delta n_s(\vec G)=\frac{1}{A}\int_A d^2r'\:
      \left\{ n_s(\vec r\: ')-n_b\right\} e^{-i\vec r\: '\cdot\vec G}.
\label{Dns}
\end{equation}
The charge neutrality of the system removes the singularity
of the Coulomb interaction for $\vec G=0$.

To complete the Hartree set of equations (\ref{V}), (\ref{H}), (\ref{VH}),
and (\ref{H0}) the electron density is written in terms of
$\{\epsilon_{\alpha}\}$ and $\{\psi_{\alpha}\}$ as
\begin{equation}
      n_s(\vec r)=\frac{1}{4\pi^2}\sum_{\alpha}\int^{\pi}_{-\pi}
      d\theta_1d\theta_2\: f(\epsilon_{\alpha}-\mu )
      |\psi_{\alpha}(\vec r)|^2,
\label{ns}
\end{equation}
where $f$ is the equilibrium Fermi distribution and $\mu$ is the
chemical potential. This set of Hartree equations has to be solved
iteratively together with the condition that the average electron
density $n_s=N_s/A$ is constant and determines the chemical potential
$\mu$. $N_s$ is the number of electrons per unit cell
with area $A$ in ${\cal B}$.

\section{Results}
The Coulomb interaction couples directly the subbands of a
particular LL and the subbands of different LL's via the
matrix elements (\ref{matrix-elements}). The matrix elements with states of
different
phase angles ($\theta_1$, $\theta_2$) do all vanish,
so that the Hartree Hamiltonian can be diagonalized for each pair
($\theta_1$, $\theta_2$) separately. However, the states for a given set of
phase angles depend on the states of all other phase angles via the Hartree
potential  (\ref{VH}) and the density $n_s(\vec r)$
(\ref{ns}). The self-consistent model described in the above section
can be used to evaluate the energy spectra $\{ \epsilon_{\alpha}\}$
and the wavefunctions $\{ \psi_{\alpha}\}$ for a wide variety of
modulation strengths $V_x$ and $V_y$ and values of the magnetic field
$B$ by selecting the appropriate size of the basis, $H$ is diagonalized
in. If only the basis states in the same LL are taken into account,
the modulation strength is irrelevant since it can be factored out of the
Hamiltonian matrix.
In this situation, corresponding to the usual discussion of Hofstadter's
butterfly on the basis of Harper's equation, a highly symmetric energy
spectrum is expected. Of course, this restricted
model is only appropriate for describing a system with
very weak modulation (as compared with the inter-Landau-level
energy spacing $\hbar\omega_c$).
In the following, parameters appropriate for GaAs have been
used, i.e. $m^*=0.067m_0$, $\kappa =12.4$, and the spin degree of
freedom has been completely neglected since the exchange force
is not accounted for in the interaction between the electrons.
The superlattice is selected to have square symmetry, $l_1=l_2=L$, and
$V_x=V_y=V_0$,
to comply with the usual experimental conditions, even though the
Ferrari-basis is applicable for a much more complicated
unit cell.\cite{Ferrari90:4598,Silberbauer92:7355,Silberbauer94:PhD}

The iteration of the Hartree equations (\ref{V}), (\ref{H}), (\ref{VH}),
and (\ref{H0}) has been accomplished by
specifying an initial electron density
that is then used to find the Hartree potential (\ref{VH}).
Subsequently the Hamiltonian (\ref{H}) is diagonalized yielding
$\{ \epsilon_{\alpha} \}$ and $\{ \psi_{\alpha} \}$ that
are in turn used to derive a new density (\ref{ns}). In order to
guarantee convergence the speed of the iteration process
is reduced by using only 10\% of the new densty and 90\%
of the density from the former iteraction cycle.
The convergence is considered completed when the
integrated rms deviations of the density from two
subsequent cycles falls below 0.01\% of $N_s$.
Clearly the convergence rate depends on modulation stength
of the external potential (\ref{V}), the filling factor $\nu$
and the temperature $T$. Commonly for the following parameters
we need between 15 and 30 cycles to attain the specified
accuracy. For the ground state properties the
evaluation of the wave functions can be avoided by
combining equations (\ref{Dns}) and (\ref{ns}) and thus write
the Fourier transformed density entirely as a function
of the analytically known matrix elements (\ref{matrix-elements}).

The energy spectrum for the case of four flux quanta $\Phi_0$ piercing
the unit cell of a Hartree interacting 2DEG periodically modulated
with $L=50\: $nm is shown in Fig.\ \ref{Fig-Ett}.
The dimensionless magnetic flux is thus, $\Phi /\Phi_0=pq=4$,
which translates into a magnetic field of $B\approx 6.617\: $T and
$\hbar\omega_c\approx  11.43\: $meV. Due to the strong modulation of
the periodic potential (\ref{V}), $V_0=4.0\: $meV, and the low
temperature $T=0.2\: $K, four separate subbands of the
lowest LB are resolved, which are centered about $E=\hbar\omega_c/2\approx
5.7\: $meV.
The electronic density is very low, $N_s=0.25$ being the average number of
 electrons per unit cell. The chemical potential $\mu =2.38\: $meV
is lying in the lowest subband which has a very slight dispersion
as function of the phase angles $\theta_1$ and $\theta_2$.
This is a manifestation of the familiar pinning effect observed
earlier in many different
systems.\cite{Wulf88:4218a,Manolescu95:1703} The relatively
weak screening (judged from the spread of the minibands $\leq 2V_0$)
is strongest in the subband where $\mu$ is situated. The next LB
is centered around $E=3\hbar\omega_c/2\approx 17.1\: $meV and has
a much weaker dispersion than the first one. The vanishing overlap
of the LB's leads to an approximate electron-hole symmetry
between the cases $N_s=0.25$ and $N_s=3.75$, since the (average)
filling factor $\nu=1$ is reached for $N_s=4$.
Note however that, e.g. for $N_s=0.25$, the energy spectrum is not symmetric
around the mean energy, in contrast to the case of the noninteracting
electrons. This approximate
symmetry can also be seen in Fig.\ \ref{Fig-Vsc} in which the self-consistent
potential $V_{sc}(\vec r)=V_H(\vec r)+V(\vec r)$ for $pq=6$ has been depicted.
Here the intersection of the shifted
chemical potential $\mu-\hbar\omega_c/2$ with $V_{sc}$
projected onto the basis of the graph is also shown.
For $N_s=0.25$ (Fig.\ \ref{Fig-Vsc}a) the 2DEG forms isolated lakes near
the minima of $V_{sc}$.
The complimentary case, $N_s=5.75$, is depicted in Fig.\ \ref{Fig-Vsc}c,
where the empty states of the first LB, ``the holes,'' are isolated
near the maxima of the potential. The transition between these two regions is
seen in  Fig.\ \ref{Fig-Vsc}b, where traditionally, i.e. in a semi classical
picture, neither the electrons nor the holes are thought to form closed
orbits. The screening is so strong for this last case (corresponding to $\nu
=1/2$) that $V_{sc}$ is almost flat. As stated earlier, the symmetry is only
approximate,  since the states of the second LB are mixed into
the states of the first one by the interaction. The potentials are
thus not identical for the complimentary cases of $N_s=0.25$ and $5.75$,
and the transition between the two regions does not happen
exactly at $N_s=3.00$. The self-consistent densities of the electrons for the
cases $N_s=0.25$ and $N_s=5.75$ is shown in Fig.\ \ref{Fig-ns},
underlining clearly the near isolation of the electrons in the minima of
$V_{sc}$ when $N_s=0.25$. In Fig.\ \ref{Fig-nsI0} we compare the
self-consistent density $n_s(\vec r)$ with the density of the
non interacting 2DEG for $N_s=3.00$. For $pq=6$ the screening is
at its maximum strength but due to the strong modulation
$V_0=4.0\: $meV the interaction only causes a moderate flattening
of the density.

The strong screening encountered in Fig.\ \ref{Fig-Vsc} for the
$pq=6$ case is even more drastic for a superlattice potential
$V(\vec r)$ with a longer period and weaker modulation as can be observed from
Fig.\ \ref{Fig-BW200} when $L=200\: $nm and $V_0=0.5\: $meV.
Here the width and the location of each subband of the lowest LB
is shown as a function of $N_s$, the location of the chemical
potential is also indicated.
Even though the temperature is quite low, $T=0.1\: $K, the Coulomb
interaction between the electrons is very effective in screening the
superlattice
potential and collapsing the subbands into an almost degenerate LL
around $N_s\sim 3$ corresponding to $\nu\sim 1/2$. For an integer filling
factor of the order of unity ($N_s\sim 6$) the screening is as expected very
weak.\cite{Wulf88:4218a} The upper panel of Fig.\ \ref{Fig-BW200} shows the
path of the chemical potential $\mu$ through the subbands as a function
of the average filling factor $\nu$ or $N_s$ when the electrons do not
interact.  The fine structure of $\mu$ leads
to a peak structure of the (dimensionlessly written)
thermodynamic density of states (TDOS)
\begin{equation}
      D_T=l^2\hbar\omega_c\left( \frac{\partial n_s}{\partial\mu}\right) _{TB},
      \quad n_s=\langle n_s(\vec r)\rangle
\label{DT}
\end{equation}
whenever $\mu$ crosses a subband. The strong screening in the case of the
interacting 2DEG (the lower panel of  Fig.\ \ref{Fig-BW200}) results in only
one peak in $D_T$,
in which no fine structure is resolved. The corresponding information is shown
for four different cases in Fig.\ \ref{Fig-BW50}, $pq=2,3,4,6$, identifiable
from the number of subbands appearing in each figure. Here the period is
short,
$L=50\: $nm, the modulation strong, $V_0=4.0\: $meV, and $T=1.0\: $K.
However only one LL is used in the calculation, so that the results apply to
the case of a weak superlattice
potential. Here the electron-hole symmetry is exact,
as is reflected in each of the four cases by the point symmetry of the
subband structure around the
chemical potential $\mu$ at half filling.
The screening is strongly dependent on the total number of states
in a lattice unit cell and
the average filling factor. It is therefore, strongest for $pq=6$ when
maximum six electrons can occupy the lowest Landau level
in the unit cell and when
the LB is half filled, $N_s=3$. The thermodynamic density of states
$D_T$, Eq. (\ref{DT}), for the corresponding cases is shown in
Fig.\ \ref{Fig-TDOS}. Clearly the strong screening in the $pq=6$
case obscures the fine structure in $D_T$ leading to only one
pronounced peak. The subbands in the other cases turn up as separate peaks
in $D_T$. The calculation has been repeated for the same parameters but
one more Landau level resulting in qualitatively the same picture.
The electron-hole symmetry is then only approximate with the largest
deviation for $pq=2,3$, but the resolution of the peaks in $D_T$ is
unchanged.

The Hofstadter butterfly should be recovered when the calculation is
performed for $N_s=0$ and only one LL, and
if the width and location of the subbands are plotted as functions of the
inverse of the number of magnetic flux quanta
through a unit cell of the lattice, i.e. of $\Phi_0/\Phi=1/(pq)$.
In addition, all energies have to be scaled according to:
$E\rightarrow (E-\hbar\omega_c/2)/V_0$. For interacting electrons
the number of electrons per unit cell $N_s$, or the average filling
factor of the LB turns out to be a new parameter controlling the
bandstructure. We therefore, present in Fig.\ \ref{Fig-Hof_Ns} the subband
structure for the
four inverse fluxes $1/(pq)=1/2,1/3,1/4,1/6$ in the four cases
$N_s=0.00,0.25,1.00,1.50$  together with
the location of the chemical potential $\mu$ (except for $N_s=0.00$, where it
is irrelevant).
For low density, say $N_s=0.25$, the subband structure is almost symmetric
around the energy zero, like in the noninteracting case, but the subbands
become quite
asymmetric for a higher density of electrons.
For the short-period superlattice
studied here, and for a low density of electrons, the essential gap structure
does survive  in the presence of interaction. The presentation in each of
the subfigures of Fig.\ \ref{Fig-Hof_Ns}  corresponds to the
experimental procedure of keeping the density of electrons fixed
but changing the magnetic field.
Another way to investigate
the screening is to keep the average filling factor $\nu$ constant but
vary the magnetic field and, thus, also the average density
of the 2DEG. Fig.\ \ref{Fig-Hof_nu} compares the subband structures
for $1/(pq)=1/2,1/3,1/4,1/6$ and $\nu =1/2$ with the complete
Hofstadter butterfly. Here the energies have been scaled with
the factor $(E-\hbar\omega_c/2)\exp\{ (\pi l/L)^2\} /V_0$ so that the results
can be directly
compared with an earlier calculation of the Hofstadter
spectrum.\cite{Pfannkuche92:12606}
The energy spectrum in the interacting case
shows an overall reduction in dispersion or bandwidth due to the
strong screening that is most effective for large flux and large number of
available states. The bandwidths of the subbands for the interacting
case has been evaluated here on a discrete lattice in the
($\theta_1$, $\theta_2$)-plane without attempting an interpolation
between the lattice points; thus, the actual bandwidths can be
larger by a small percentage of the widths shown. This effect
explains why the bands for the even denominators in Fig.\ \ref{Fig-Hof_Ns}
for $N_s=0$ are not touching.

\section{Summary}
In an interacting 2DEG subject to a superlattice potential and a homogeneous
perpendicular magnetic field not only the magnetic flux through a
unit cell but also the density of electrons determines the
complicated splitting of the Landau bands into subbands.
We have shown that in the Hartree approximation the essential
gap structure of the energy spectrum remains,
although the screening leads to a quenching of the Hofstadter butterfly
at small values of the inverse flux (see Fig.~8). The symmetry of the
butterfly is also lowered by
the coupling to higher LL's due to increased strength of
the periodic potential, as has already been discussed by other
authors.\cite{Kuhn93:8225,Petschel93:239}
For periods around $L=200\: $nm (currently attainable in
experiments on superlattices) the 2DEG can effectively screen the
periodic potential even for a very low density $n_s$ so that only one
or two Landau bands are partly occupied. Only at shorter
lattice constants ($L<100\: $nm) and thus much higher magnetic fields
we can, on the basis of our Hartree calculation, expect the subband
structure to be resolvable in experiments, when $n_s$
is maintained low enough. These predictions are made on
the basis of the calculated
energy spectrum and the structure observed in the thermodynamic density
of states $D_T$, see Eq. (\ref{DT}). The transport coefficients
do not depend on $D_T$ in any simple way,
and may be more sensitive to the subband structure of the energy
spectrum.\cite{Pfannkuche92:12606}
In addition, the collision broadening due to impurities and
inhomogeneities has to be considered, especially, when the electronic
density is very low.\cite{Pfannkuche92:12606,Vasilopoulos91:177,Wulf93:6566}

It is to be expected that
the inclusion of exchange and correlation effects may lead to more dramatic
changes in the
energy spectrum than the direct interaction considered in the
Hartree approximation. One might even expect rearrangement of the
bands or the occurence of
spin-density waves.\cite{Gudmundsson94:92,Manolescu95:1703,Pfannkuche93:2244}
A reliable procedure for the treatment of exchange and correlation
effects in the 2DEG in a strong magnetic field and a lateral
superlattice is, however, at present not available. Therefore we have
restricted our investigation of the electron-electron interaction effects
by considering the Hartree approximation, which is expected to yield a
qualitatively
correct description of the dominant screening effects in inhomogeneous
electron systems.

In order to keep the computational efforts in reasonable limits, we have
restricted our calculations to a few characteristic values of the magnetic
field. Our results indicate in which manner screening effects will change
the overall appearence of
the Hofstadter energy spectrum. Unfortunately, there is no
simple way to extrapolate from these special
values to arbitrary rational numbers of flux quanta per unit cell. For a
more detailed comparison with the energy spectrum of the noninteracting
case, further studies using a much denser set of flux values are
desirable. However, the answer of the interesting     question,
whether the self-similarity of the Hofstadter butterfly survives in the
presence of the mutual Coulomb interactions between the electrons, at
least within the Hartree approximation, demands an amount of computational
work which seems at present unfeasible. On the other hand, from the
experimental point of view these details of the energy spectrum seem not
to be accessible anyway. The challenge for the near future is to
resolve the most prominent gaps of the energy spectrum experimentally.

\acknowledgements
We are grateful to Behnam Farid for a critical reading of the manuscript.
This research was supported in part by the Icelandic Natural Science
Foundation, the University of Iceland Research Fund,
and a NATO collaborative research Grant No. CRG 921204.
\bibliographystyle{prsty}

\newpage
\begin{figure}
\caption{The dispersion of the four minibands of the lowest Landau level in
         the energy spectrum of the Hartree interacting 2DEG
         as a function of the phases $\theta_1$ and $\theta_2$
         when four flux quanta of the magnetic field pierce
         a rectangular unit lattice cell ($p=q=2$). $L_x=L_y=L=50\: $nm,
         $T=0.2\:$ K, $V_x=V_y=V_0=4.0\: $meV, $N_s=0.25$, and
         $\mu =2.38\: $meV. Two Landau levels are included in the
         calculation.
}
\label{Fig-Ett}
\end{figure}
\begin{figure}
\caption{The self-consistent potential
         $V_{sc}(\vec r)=V_H(\vec r)+V(\vec r)$ in a unit
         lattice cell for (a) $N_s=0.25$, (b) $N_s=3.00$, and (c)
         $N_s=5.75$. The intersection of the
         chemical potential $\mu-\hbar\omega_c/2$
         and $V_{sc}$ is projected on the
         base and $\mu-\hbar\omega_c/2$ is shown in the
         upper right corner of each
         subfigure in meV. $p=2$, $q=3$,
         $L=50\: $nm, $T=1.0\: $K, $V_0=4.0\: $meV.
         Two Landau levels are included in the calculation.
}
\label{Fig-Vsc}
\end{figure}
\begin{figure}
\caption{The electron density $n_s(\vec r)$ ($nm^{-2}$)
         in a unit lattice cell
         for (a) $N_s=0.25$, and (b) $N_s=5.75$.
         Contours of $n_s(\vec r)$ are shown at the base.
         $p=2$, $q=3$, $L=50\: $nm, $T=1.0\: $K, $V_0=4.0\: $meV.
         Two Landau levels are included in the calculation.
}
\label{Fig-ns}
\end{figure}
\begin{figure}
\caption{The electron density $n_s(\vec r)$ ($nm^{-2}$)
         in a unit lattice cell for $N_s=3.00$.
         (a) Non interacting, and (b) Hartree interacting 2DEG.
         Contours of $n_s(\vec r)$ are shown at the base.
         Same scales are used in both cases.
         $p=2$, $q=3$, $L=50\: $nm, $T=1.0\: $K, $V_0=4.0\: $meV.
         Two Landau levels are included in the calculation.
}
\label{Fig-nsI0}
\end{figure}
\begin{figure}
\caption{The bandwidth of the minibands of the lowest Landau
         band as a function of the number of electrons per
         unit cell $N_s$ for a noninteracting 2DEG (upper)
         and a Hartree interacting 2DEG (lower).
         The chemical potential $\mu$ is noted by a continuous curve.
         $L=200\: $nm, $T=0.1\: $K, $V_0=0.5\: $meV, $p=2$,
         and $q=3$. Two Landau levels are included in the calculation.
}
\label{Fig-BW200}
\end{figure}
\begin{figure}
\caption{The bandwidth of the minibands of the lowest Landau
         band as a function of the number of electrons per
         unit cell $N_s$ for a Hartree interacting 2DEG and
         ($p=1$, $q=2$) (upper left), ($p=1$, $q=3$) (upper right),
         ($p=2$, $q=2$) (lower left), and ($p=2$, $q=3$) (lower right).
         The chemical potential $\mu$ is noted by a continuous curve.
         $L=50\: $nm, $T=1.0\: $K, $V_0=4.0\: $meV.
         One Landau level is included in the calculation.
}
\label{Fig-BW50}
\end{figure}
\begin{figure}
\caption{The thermodynamic density of states
         $l^2\hbar\omega_c(\partial n_s/\partial\mu)_{TB}$
         as a  function of the dimensionless inverse magnetic flux through
         a unit cell $1/pq$ and $N_s$.
         $L=50\: $nm, $T=1.0\: $K, $V_0=4.0\: $meV.
         One Landau level is included in the Hartree calculation.
}
\label{Fig-TDOS}
\end{figure}
\begin{figure}
\caption{The scaled bandwidth of the subbands of the lowest Landau band
         $(E-\hbar\omega_c/2)/V_0$ as function
         of the dimensionless inverse magnetic flux $1/pq$ for $N_s=0.00$
         (upper left), $N_s=0.25$ (upper right), $N_s=1.00$
         (lower left), and $N_s=1.50$ (lower right).
         The chemical potential $\mu$ is noted by crosses.
         $L=50\: $nm, $T=1.0\: $K, $V_0=4.0\: $meV.
         One Landau level is included in the Hartree calculation.
}
\label{Fig-Hof_Ns}
\end{figure}
\begin{figure}
\caption{The scaled bandwidth $(E-\hbar\omega_c/2)\exp\{ (\pi l/L)^2\} /V_0$
         of the subbands of the lowest Landau band as function
         of the dimensionless inverse magnetic flux $1/pq$ for the Hartree
         interacting 2DEG with $\nu=1/2$ (left), and for the
         noninteracting 2DEG (right), the Hofstadter butterfly.
         One Landau level is included in the Hartree calculation.
         In the left subfigure the 6 LB's for the case $pq=6$ can
         not all be resolved due to vanishing band gaps.
         The parameters used for the left subfigure are:
         $L=50\: $nm, $T=1.0\: $K, $V_0=4.0\: $meV.
}
\label{Fig-Hof_nu}
\end{figure}
\end{document}